# Combine Statistical Thinking With Open Scientific Practice: A Protocol of a Bayesian Research Project

Alexandra Sarafoglou[1], Anna van der Heijden[1,2], Tim Draws[3], Joran Cornelisse[1,4], Eric-Jan Wagenmakers[1], and Maarten Marsman[1]

[1] Department of Psychology, University of Amsterdam, The Netherlands

[2] Currently at De Groene Lobby, Haarlem, The Netherlands

[3] Faculty of Engineering, Mathematics and Computer Science, Delft University of Technology, The Netherlands

[4] Currently at SocialDatabase, Amsterdam, The Netherlands

Abstract

Current developments in the statistics community suggest that modern statistics education should be structured holistically, that is, by allowing students to work with real data and to answer concrete statistical questions, but also by educating them about alternative frameworks, such as Bayesian inference. In this article, we describe how we incorporated such a holistic structure in a Bayesian research project on ordered binomial probabilities. The project was conducted with a group of three undergraduate psychology students who had basic knowledge of Bayesian statistics and programming, but lacked formal mathematical training. The research project aimed to (1) convey the basic mathematical concepts of Bayesian inference; (2) have students experience the entire empirical cycle including collection, analysis, and interpretation of data and (3) teach students open science practices.

## 1. Introduction

The curriculum guidelines of the American Statistical Association (ASA) argue that statistics education in undergraduate programs should not be primarily focused on teaching statistical methods and mathematical foundations, but also emphasize scientific practice, that is, study design, data collection, programming skills, and data analysis (American Statistical Association, 2014; Horton & Hardin, 2015; Wasserstein et al., 2016). In general, students should learn to "think with and about data" Cobb, 2015, p. 267 and thus develop a holistic understanding of statistics (Horton & Hardin, 2015).

This holistic understanding of statistics also includes learning and understanding alternatives to classical inference based on *p*-values. Bayesian inference is becoming increasingly popular and its adoption has been advocated for both scientific practice (Wasserstein et al., 2016) and statistics education (Cobb, 2015). Recent examples for undergraduate courses and in-class demonstrations on Bayesian methods that require only little or no mathematical or statistical training are described in Witmer (2017; on teaching Markov chain Monte Carlo methods), Rouder and Morey (2018; on teaching Bayes' rule), and van Doorn, Matzke and Wagenmakers (2020; on teaching the key concepts of Bayesian inference). However, little attention has been paid to the design and structure of Bayesian research projects that can be conducted with a small group of students, for instance, in the context of a thesis or internship project, or a seminar. These formats, as opposed to standard courses, allow for more extensive research projects, since supervisors can offer individual support for students, and can dedicate more time to the execution of the project.

We believe that a research project on Bayesian inference should take advantage of the rather long project duration and the small group size by introducing students in detail to the theoretical and practical aspects of Bayesian inference. Theoretical aspects of Bayesian inference entail that by the end of the project students should feel comfortable with the

---

Correspondence concerning this article should be addressed to: Alexandra Sarafoglou, Nieuwe Achtergracht 129B, 1001 NK Amsterdam, The Netherlands, E-mail: alexandra.sarafoglou@gmail.com.

standard terminology, be able to understand how to assign a prior distribution, specify a likelihood function, derive a posterior distribution, and compute a marginal likelihood. The practical aspects entail that students should be able to apply their theoretical knowledge to address a concrete research question, and experience the entire empirical cycle, including study planning, preregistration, data collection and analysis, and interpretation of the results. These teaching goals resulted in three guiding principles for structuring the project, listed below.

The first principle is to introduce students to the mathematics underlying Bayesian statistics. In our own teaching of Bayesian methods in undergraduate psychology courses, we usually hide the mathematics and instead aim to provide students with an intuition about how Bayesians use distributions to quantify uncertainty about model parameters and hypotheses. This approach helps students interpret posterior distributions, credible intervals, and Bayes factors (for a gentle technical introduction to Bayesian inference without mathematical derivations see Etz and Vandekerckhove, 2018). However, for students who want to specialize in research methods and statistics it is important to go beyond an intuitive understanding and be introduced to the mathematics behind these key concepts. Without the mathematical foundations, students will find the statistical literature difficult to understand.

The second principle is to let students experience scientific practice. In line with the ASA guidelines on statistics education (American Statistical Association, 2014), we believe that students learn most when they are given the opportunity to gain hands-on experience on how to apply the methods taught to a real data example. We therefore set up a Bayesian replication study that demonstrates a series of Bayesian benefits. For instance, in contrast to frequentist analyses, the Bayesian framework allows students (1) to discriminate between "absence of evidence" and "evidence of absence" of the effect in the replication study (Dienes, 2014; Keysers et al., 2020; Verhagen & Wagenmakers, 2014); (2) to experience Bayesian learning by incorporating prior knowledge –such as data from previous experiments– to construct a more informative test (Verhagen & Wagenmakers, 2014); (3) to monitor evidence as the data accumulates (Rouder, 2014). In addition, it allows students to learn how conclusions from significant *p*-values differ from conclusions drawn from Bayes factors by conducting a Bayesian reanalysis of the results of the original experiment.[1]

The third and final principle is to convey open science practices. Reproducibility and replicability are core scientific values, but yet psychological science is currently facing a crisis of confidence as a disappointing proportion of key findings appear to be reproducible (Baker, 2016; Camerer et al., 2018; Klein et al., 2014; Klein et al., 2018; Nature Publishing Group, 2016; Open Science Collaboration, 2015; Pashler & Wagenmakers, 2012). To a large extent, the low rate of reproducible findings can be attributed to the great flexibility in data analysis in combination with selective reporting of significant results (Simmons et al., 2011), the high prevalence of questionable research practices (John et al., 2012), the reluctance to conduct direct replication studies (Pashler & Harris, 2012; Schmidt, 2009), and the poor availability of research data (Houtkoop et al., 2018; Wicherts et al., 2006; for a special issue on data sharing see Simons, 2018). To address these problems, psychological science today relies on numerous open scientific practices, such as preregistration and Registered Reports, large-scale collaborations, and sharing of data, materials, and code (e.g., Chambers, 2013; Chambers & Tzavella, 2021; Kidwell et al., 2016; Morey et al., 2016; Moshontz et al., 2018; Nosek et al., 2015). However, to truly integrate these practices into the research culture, it is necessary to introduce the principles of open science to students at an early stage (Chopik et al., 2018; Funder et al., 2014; Morling & Calin-Jageman, 2020; Munafò et al., 2017; Sarafoglou et al., 2020). Since thesis projects often require detailed design and analysis plans we view them as a good opportunity for supervisors to teach them both the philosophy behind open science and the practical skills needed to apply open science practices. Therefore, we set up a preregistered replication study, and have students publish the analysis code, and share the data and materials on the Open Science Framework (OSF; Center for Open Science, 2021).

[1] We refer the interested reader to Wagenmakers (2007) and Wagenmakers et al. (2018) for a more detailed discussion on the benefits of Bayesian inference.

The purpose of this article is to share our experiences on designing and supervising a Bayesian thesis project for undergraduate psychology students. Lecturers who intend to offer a Bayesian research project for a small group of students which emphasizes mathematical training as well as practical experience with real data might find helpful advice on what focal points to set when planning their project. In addition, the described project can serve as illustrative example in a classroom setting, to teach students Bayesian learning, and a simple method to evaluate ordinal expectations. In the following, we will describe the course structure, the theoretical and the practical part of the project in more detail.

## 2. Supplemental Material

Interested readers can visit our OSF project folder (https://osf.io/zfhbc/) to access the following information: the study preregistration, the analysis code, all data and materials, and the student evaluations. Furthermore, it contains the results of the Bayesian reanalysis of the original studies and the formal description of the mathematical model for multiple independent binomial probabilities.

## 3. Project Overview

Here we describe the thesis project titled “A Bayesian View on ‘Science versus the Stars’: Bayes factor analysis for ordered binomial probabilities” at the University of Amsterdam. The topic of the thesis project was the Bayesian analysis of ordinal expectations of multiple binomial probabilities. We chose this topic due to both its relevance in the psychological literature and the simplicity of the statistical model. Ordinal expectations of binomial probabilities are common in the area of psychometrics and theories on rational decision making (see e.g., Cavagnaro & Davis-Stober, 2014; Davis-Stober, 2009; Guo & Regenwetter, 2014; Haaf et al., 2020; Heck & Davis-Stober, 2019; Myung et al., 2005; Regenwetter et al., 2018; Regenwetter et al., 2011; Tijmstra et al., 2015). For instance, a psychometrician who evaluates whether a test for cognitive performance can be measured on an interval scale needs to test the assumption that the probability to solve a given item is non-decreasing for the ability of a person. One argument to use Bayesian methods for these problems is that we can easily incorporate ordinal expectations of the binomial probabilities in the respective prior distributions (Klugkist et al., 2010). This makes the corresponding statistical model particularly simple and enables students to derive the method even without formal mathematical training.

During the theoretical part of the project students familiarized themselves with the computation of Bayes factors for ordered binomial probabilities using the encompassing prior method (Klugkist et al., 2005). During the practical part of the project the students applied the methods in practice by conducting a preregistered reanalysis and replication study.

## 3.1. Course Structure

The full thesis project–starting from the first introductory lesson to submission of the research report–took 16 weeks. A weekly overview of the research project is provided in Table 1. Our students had to hand-in two writing assignments, create the preregistration of the empirical study, and write the final report. On average the students worked 22−23 hours per week on the project for which they were rewarded with 12 ECTS credits. The following section describes these components in more detail.

## 3.2. Supervision

During the theoretical part of the project we supervised the students intensively; we had weekly group meetings that were structured as lectures, we gave students two writing assignments, and we reviewed and discussed these assignments with each student individually. During the practical part, the students then primarily worked independently with little need for supervision. The weekly group meetings were replaced by individual contact hours that gave students the opportunity to discuss details of their report.

**Table 1**

*A week-by-week overview of our project "A Bayesian view on science versus the stars: Bayes factor analysis for ordered binomial probabilites".*

| Week | Goal | Activities |
|---|---|---|
| 1 | Reiterating knowledge | Bayesian parameter estimation and hypothesis testing for the beta-binomial model<br>Write methods section of research report |
| 2 | Establishing knowledge | Generalize concepts to multiple binomials<br>Write methods section of research report |
| 3 | Establishing knowledge | Derive and apply Savage-Dickey density ratio<br>Write methods section of research report |
| 4 | Establishing knowledge | Derive and apply encompassing prior approach<br>Bayesian reanalysis of Carlson (1985) and Wyman and Vyse (2008)<br>Write introduction of research report |
| 5 | Writing | Finalize the methods section of the research report;<br>Write introduction of research report |
| 6–7 | Preregister study | Plan replication study<br>Create preregistration document |
| 8 | Preregister study | Print all necessary documents, prepare data collection (e.g, book lab)<br>Finalize preregistration |
| 9–10 | Data collection 1 | Participants fill out NEO-FFI and report date and place of birth |
| 11 | Create study materials | Generate personality descriptions<br>Prepare follow-up data collection |
| 12–13 | Data collection 2 | Participants perform choice task |
| 14 | Analyzing data | Analyze data and upload the dataset to the OSF<br>Write results section of research report |
| 15–16 | Finalizing project | Finalize research report<br>Prepare 20-minute presentation |

### 3.3. Writing Assignments

We dedicated the first four sessions at the beginning of the project to the theoretical concepts of Bayesian inference. During these four weeks, students wrote two short 1 −2 page reports explaining Bayesian parameter estimation and hypothesis testing. The first report concerned Bayesian inference for one binomial probability. In the second report, they had to generalize these concepts to multiple binomial probabilities. Students could incorporate these reports as part of the methods section in their final report.

### 3.4. Preregistration

Our students had three weeks to create the preregistration document. Since our students answered the same research question, we let them create the preregistration together. The preregistration featured the following components: the study design; the sampling plan, sampling plan rationale, and stopping rule for data collection; exclusion criteria; the description of the materials and procedure; the research question and hypotheses (including the expected direction of the effect); details on the statistical model and analysis plan, including specifications for prior distributions, number of samples drawn, inference criteria, and handling of missing data.

### 3.5. Grading Criteria

For the most part, grading was based on the individual research reports. We assessed whether students were able to (1) justify the proposed research question and methods; (2) describe the Bayesian concepts accurately by using the specific terminology; (3) discuss and interpret the results correctly; and (4) adopt a scientific writing style. In addition, students could receive a pass or fail both on their final presentation and on their learning progress. The writing assignments and the preregistration were not graded.

## 4. The Theoretical Part: Bayesian Parameter Estimation and Hypothesis Testing For Multiple Binomial Probabilities

The goal for the theoretical part of the project was to teach students when and how the encompassing prior approach is used, and how it is derived. To ease the students into this topic, we asked them to reiterate the basic mathematical concepts in Bayesian inference by means of one binomial success probability, that is, Bayesian parameter estimation (including Bayes' rule, the prior distribution, the likelihood function, marginal likelihood, and the posterior distribution) and Bayesian hypothesis testing (including prior model odds, the Bayes factor, and posterior model odds). Subsequently, students had to generalize these concepts to multiple binomial success probabilities.

# 5. The Practical Part: Reanalysis and Replication of Wyman and Vyse

We searched for empirical studies which involved hypotheses about the ordering of multiple binomial probabilities. The study by Wyman and Vyse (2008) is a suitable candidate for a replication study, for several reasons. First, the study had an engaging research question, that is, whether the accuracy of psychological personality descriptions is similar to the accuracy of astrological natal charts. Second, the dependent variables in Wyman and Vyses' study allowed for the formulation of an ordinal expectation. Third, replicating the study did not require knowledge about sophisticated concepts such as item response theory (Birnbaum, 1968; Rasch, 1960). Fourth, the experimental setup for the study was straightforward which made the planning and execution of a preregistered replication study feasible for our time frame. Note that the study by Wyman and Vyse is itself a conceptual replication of a study conducted by Carlson (1985). We chose to replicate the study by Wyman and Vyse (2008), however, since the authors had a clearer setup and material that was easier to reproduce. We aimed to replicate the study by Wyman and Vyse (2008) as closely as possible which meant that we adapted the original research design with only a few practical changes.

## 5.1. Methods

### *5.1.1. Sampling plan*

We preregistered to collect data from 50 participants or stop data collection by June 1st, 2018. The target sample size was based on the number of participants in the original studies which was 56 participants in Carlson (1985) and 52 participants in Wyman and Vyse (2008). Unfortunately, we were not able to reach the preregistered target sample before our testing period ended. We were only able to recruit 31 participants. Of those, 2 participants did not attend the second meeting, leaving us with a final sample of 29 participants.

### 5.1.2. *Materials*

In their study, Wyman and Vyse used the NEO Five Factor inventory (NEO-FFI, Costa and McCrae, 1985, 1992) to create psychological personality descriptions and the software Astrolabe (Astrolabe Inc, 2018) to create astrological natal charts for each participant. Then, an experimenter gave each participant their own psychological personality description and a psychological personality description belonging to another participant. The participant was then asked to decide which of the two personality descriptions was their own. This procedure was then repeated for the astrological personality description.

### 5.1.3. *Procedure*

The research design required two testing periods that were one week long and approximately two weeks apart. During the first testing period, the students assessed participants with the NEO-FFI personality inventory and collected information, that is, date and place of birth, that allowed them to create astrological natal charts for each participant with the free version of the software used by the original authors. In the second testing period the participants had to perform a simple choice task; they were asked to identify both their own psychological personality description and their astrological natal chart out of two descriptions each (i.e., a chance level of 50%).

### 5.1.4. *Hypotheses*

For the replication, our students took into account the direction of the original results and thus tested the ordinal hypothesis $H_r$ that the success probability for psychological personality descriptions is *higher* than that for astrological personality descriptions. This hypothesis was then tested against a point-null hypothesis $H_0$ that both success probabilities are equal to chance. Furthermore, as will be explained in the next section, the calculation of the Bayes factor required another hypothesis –referred to as the encompassing hypothesis $H_e$– that both success probabilities can vary freely.

#### 5.1.5. *Analysis Plan*

The students assigned a beta prior distribution to the model parameters and used the data from Wyman and Vyse (2008) to inform their prior beliefs. Specifically, based on Wyman and Vyse's data, the students assigned a Beta(42, 12) prior distribution to the probability of correctly identifying one's own psychological personality description and a Beta(25, 29) prior distribution to the success probability of correctly identifying one's own astrological personality description. That is, the prior for psychological personality descriptions favors success probabilities well above chance level while for astrological personality descriptions success probabilities at chance level are favored, with medians and 95% credible intervals of 0.77 [0.65, 0.87] and 0.46 [0.34, 0.59], respectively. To compare $H_0$ versus $H_r$, the students first had to take a two-step approach. First, they needed to compute the Bayes factor between $H_0$ and $H_e$, denoted as $\text{BF}_{0e}$, using the Savage-Dickey density ratio (Dickey & Lientz, 1970; Wagenmakers et al., 2010) and the Bayes factor between $H_r$ and $H_e$, denoted as $\text{BF}_{re}$, using the encompassing prior approach (Klugkist et al., 2005). The students then obtained $\text{BF}_{r0}$ through transitivity, that is: $\text{BF}_{r0} = \text{BF}_{re} \times \text{BF}_{e0}$. A detailed description of the statistical model is available in the online appendix.

### 5.2. Results of the replication study

In our replication study, out of 29 participants, 25 correctly identified their own psychological personality description and 18 participants correctly identified their own astrological personality description (see Table 2). Given our data and the prior knowledge provided by the Wyman and Vyse study, the result suggests extreme evidence (i.e., $\text{BF}_{r0} = 1884$) in favor of the hypothesis that people recognize their psychological personality description more reliably than their astrological personality description.

## 6. Considerations for Lecturers

Our experience with this project suggests that three considerations warrant special attention. First, lectures should be aware of the prior knowledge of their students. The

**Table 2**

*Data from the current research project, as well as from Wyman and Vyse (2008) and Carlson (1985), where $x_{psy}$ and $x_{astro}$ denote the number of participants who correctly identified their psychological personality description and their astrological personality description, respectively, $n_{psy}$ and $n_{astro}$ denote the respective total number of observations, and $\hat{\theta}_{psy}$ and $\hat{\theta}_{astro}$ denote the sample proportions of correctly identifying one's own personality description.*

| | Data | | | | | | |
|---|---|---|---|---|---|---|---|
| Study | $x_{\text{psy}}$ | $n_{\text{psy}}$ | $\hat{\theta}_{\text{psy}}$ | $x_{\text{astro}}$ | $n_{\text{astro}}$ | $\hat{\theta}_{\text{astro}}$ | Chance level |
| Current project | 25 | 29 | 0.86 | 18 | 29 | 0.62 | 0.50 |
| Wyman and Vyse (2008) | 41 | 52 | 0.79 | 24 | 52 | 0.46 | 0.50 |
| Carlson (1985) | 25 | 56 | 0.45 | 28 | 83 | 0.38 | 0.33 |

described project was designed for students who have some knowledge of Bayesian statistics, but also a basic background in the programming language R (R Core Team, 2021). However, despite their familiarity with the key concepts of Bayesian inference, our students found the mathematical parts of the project particularly challenging. Therefore, we recommend lectures to allow enough time to reiterate necessary mathematical components.

Second, when students are required to independently draft the preregistration we recommend the use of preregistration templates. For instance, the OSF offers preregistration templates for standard empirical research, but also replication studies (see https://osf.io/zab38/wiki/home/ for an overview of all preregistration forms). The transparency checklist by Aczel et al. (2020) is another highly accessible tool which covers the most important aspects for achieving transparency and openness in preregistrations and manuscripts.

Finally, in the described project, we based our target sample size on the number of participants in the original studies. Alternatively, lecturers could based their target sample size on a Bayesian design analysis (Stefan et al., 2019). A Bayesian design analysis is considered the Bayesian version of a frequentist power analysis and allows researchers to determine the minimum number of participants needed to achieve compelling evidence either in favor or against the hypothesis. Lecturers could also choose to do sequential testing, that is, monitor the evidence as the data accumulates and stop data collection as soon as the evidence is sufficiently compelling e.g., Rouder, 2014.

## 7. Summary

The discussed research project allowed students to learn a relevant Bayesian method to compute Bayes factors for ordinal expectations (i.e., the encompassing prior approach), and increase their understanding of the underlying mathematical concepts of Bayesian inference. We believe that this learning success was primarily due to the simplicity of the discussed statistical model which enabled the students to formulate the likelihood function, assign a prior distribution, derive the posterior distribution, and understand the encompassing prior approach even without strong mathematical background.

In addition, students gained practical experience through designing and conducting a reanalysis and replication study. Through this experience the students learned the advantages of Bayesian statistics in the context of replication research, for instance, by being able to quantify evidence for the absence of the predicted effect, but also by incorporating prior knowledge into their analyses and hence draw more informed decisions. In addition, the project gave students the opportunity to practice open research practices by letting them preregister their study, that is, create an analysis plan prior to data collection, and share their data, materials, and code. The confrontation with real data challenged the students to think in broader terms, that is, by discovering how different methods (i.e., the Savage-Dickey density ratio and the encompassing prior approach) can be utilized to answer specific research questions.

We believe that a research project is an ideal opportunity to integrate the theory and mathematics of Bayesian inference with hands-on experience, and confront students with all aspects of the empirical cycle. This experience gives students valuable insights into scientific practice, and equips them with problem solving skills that are necessary when they pursue their careers as psychological researchers and methodologists.

## Acknowledgements

This research was supported by a Netherlands Organisation for Scientific Research (NWO) grant to AS (406-17-568), a Veni grant from the NWO to MM (451-17-017), as well as a Vici grant from the NWO to EJW (016.Vici.170.083). At the time the research was conducted, the authors AvdH, TD and JC were affiliated with the University of Amsterdam, the Netherlands.

## Author Contact Information

Correspondence concerning this article should be addressed to: Alexandra Sarafoglou, Department of Psychology, PO Box 15906, 1001 NK Amsterdam, The Netherlands, E-mail: alexandra.sarafoglou@gmail.com.

## References

Aczel, B., Szaszi, B., Sarafoglou, A., Kekecs, Z., Kucharský, Š., Benjamin, D., Chambers, C. D., Fisher, A., Gelman, A., Gernsbacher, M. A., Ioannidis, J. P., Johnson, E., Jonas, K., Kousta, S., Lilienfeld, S. O., Lindsay, D. S., Morey, C. C., Munafò, M., Newell, B. R., . . . Wagenmakers, E.-J. (2020). A consensus-based transparency checklist. *Nature Human Behaviour*, *4*, 4–6.

American Statistical Association. (2014). Curriculum guidelines for undergraduate programs in statistical science.

Astrolabe Inc. (2018). Astrolabe [Computer software].

Baker, M. (2016). 1,500 scientists lift the lid on reproducibility. *Nature News*, *533*, 452.

Birnbaum, A. (1968). Some latent trait models and their use in inferring an examinee's ability. In F. Lord & M. Novick (Eds.), *Statistical theories of mental test scores* (pp. 395–479). Addison-Wesley.

Camerer, C. F., Dreber, A., Holzmeister, F., Ho, T.-.-H., Huber, J., Johannesson, M., Kirchler, M., Nave, G., Nosek, B., Pfeiffer, T., Altmejd, A., Buttrick, N., Chan, T., Chen, Y., Forsell, E., Heikensten, E., Hummer, L., Imai, T., Isaksson, S., . . . Wu, H. (2018). Evaluating replicability of social science experiments in *Nature* and *Science*. *Nature Human Behaviour*, *2*, 637–644.

Carlson, S. (1985). A double-blind test of astrology. *Nature*, *318*, 419–425.

Cavagnaro, D. R., & Davis-Stober, C. P. (2014). Transitive in our preferences, but transitive in different ways: An analysis of choice variability. *Decision*, *1*, 102–122.

Center for Open Science. (2021). Open Science Framework.

Chambers, C. D. (2013). Registered Reports: A new publishing initiative at *Cortex*. *Cortex*, *49*, 609–610.

Chambers, C. D., & Tzavella, L. (2021). The past, present and future of Registered Reports. *Nature Human Behaviour*, 1–14.

Chopik, W., Bremner, R., Defever, A., & Keller, V. (2018). How (and whether) to teach undergraduates about the replication crisis in psychological science. *Teaching of Psychology*, *45*, 158–163.

Cobb, G. (2015). Mere renovation is too little too late: We need to rethink our undergraduate curriculum from the ground up. *The American Statistician*, *69*, 266–282.

Costa, P. T., & McCrae, R. R. (1985). *The NEO personality inventory*. Psychological Assessment Resources.

Costa, P. T., & McCrae, R. R. (1992). *NEO-PI-R and NEO-FFI professional manual* (Vol. 38).

Davis-Stober, C. P. (2009). Analysis of multinomial models under inequality constraints: Applications to measurement theory. *Journal of Mathematical Psychology*, *53*, 1–13.

Dickey, J. M., & Lientz, B. (1970). The weighted likelihood ratio, sharp hypotheses about chances, the order of a Markov chain. *The Annals of Mathematical Statistics*, *41*, 214–226.

Dienes, Z. (2014). Using Bayes to get the most out of non-significant results. *Frontiers in psychology*, *5*, 781.

Etz, A., & Vandekerckhove, J. (2018). Introduction to Bayesian inference for psychology. *Psychonomic Bulletin & Review*, *25*, 5–34.

Funder, D. C., Levine, J. M., Mackie, D. M., Morf, C. C., Sansone, C., Vazire, S., & West, S. G. (2014). Improving the dependability of research in personality and social psychology: Recommendations for research and educational practice. *Personality and Social Psychology Review*, *18*, 3–12.

Guo, Y., & Regenwetter, M. (2014). Quantitative tests of the perceived relative argument model: Comment on Loomes (2010). *Psychological Review*, *121*, 696–705.

Haaf, J. M., Merkle, E. C., & Rouder, J. N. (2020). Do items order? the psychology in IRT models. *Journal of Mathematical Psychology*, *98*. https://doi.org/https://doi.org/10.1016/j.jmp.2020.102398

Heck, D. W., & Davis-Stober, C. P. (2019). Multinomial models with linear inequality constraints: Overview and improvements of computational methods for Bayesian inference. *Journal of Mathematical Psychology*, *91*, 70–87.

Horton, N. J., & Hardin, J. S. (2015). Teaching the next generation of statistics students to "Think with Data": Special issue on statistics and the undergraduate curriculum. *The American Statistician*, *69*, 259–265.

Houtkoop, B. L., Chambers, C., Macleod, M., Bishop, D. V., Nichols, T. E., & Wagenmakers, E.-J. (2018). Data sharing in psychology: A survey on barriers and preconditions. *Advances in Methods and Practices in Psychological Science*, *1*, 70–85.

John, L., Loewenstein, G., & Prelec, D. (2012). Measuring the prevalence of questionable research practices with incentives for truth telling. *Psychological Science*, *23*, 524–532.

Keysers, C., Gazzola, V., & Wagenmakers, E.-J. (2020). Using Bayes factor hypothesis testing in neuroscience to establish evidence of absence. *Nature Neuroscience*, *23*, 788–799.

Kidwell, M. C., Lazarević, L. B., Baranski, E., Hardwicke, T. E., Piechowski, S., Falkenberg, L.-.-S., Kennett, C., Slowik, A., Sonnleitner, C., Hess–Holden, C., Errington, T. M., Fiedler, S., & Nosek, B. A. (2016). Badges to acknowledge open practices: A simple, low cost, effective method for increasing transparency. *PLOS Biology*, *14*, e1002456.

Klein, R., Ratliff, K., Vianello, M., Adams, R. B., Jr., Bahník, Š., Bernstein, M., Bocian, K., Brandt, M., Brooks, B., Brumbaugh, C., Cemalcilar, Z., Chandler, J., Cheong, W., Davis, W., Devos, T., Eisner, M., Frankowska, N., Furrow, D., Galliani, E. M., . . . Nosek, B. (2014). Investigating variation in replicability: A “Many Labs” replication project. *Social Psychology*, *45*, 142–152.

Klein, R., Vianello, M., Hasselman, F., Adams, B., Adams, R., Alper, S., Aveyard, M., Axt, J., Bahník, S., Batra, R., Berkics, M., Bernstein, M., Berry, D., Bialobrzeska, O., Binan, E., Bocian, K., Brandt, M., Busching, R., Redei, A., . . . Nosek, B. (2018). Many Labs 2: Investigating variation in replicability across sample and setting. *Advances in Methods and Practices in Psychological Science*, *1*, 443–490.

Klugkist, I., Kato, B., & Hoijtink, H. (2005). Bayesian model selection using encompassing priors. *Statistica Neerlandica*, *59*, 57–69.

Klugkist, I., Laudy, O., & Hoijtink, H. (2010). Bayesian evaluation of inequality and equality constrained hypotheses for contingency tables. *Psychological Methods*, *15*, 281–299.

Morey, R. D., Chambers, C. D., Etchells, P. J., Harris, C. R., Hoekstra, R., Lakens, D., Lewandowsky, S., Coker Morey, C., Newman, D. P., Schönbrodt, F. D., Vanpaemel, W., Wagenmakers, E.-J., & Zwaan, R. A. (2016). The peer reviewers’ openness initiative: Incentivizing open research practices through peer review. *Royal Society Open Science*, *3*, 150547.

Morling, B., & Calin-Jageman, R. J. (2020). What psychology teachers should know about open science and the new statistics. *Teaching of Psychology*, *47*, 169–179.

Moshontz, H., Campbell, L., Ebersole, C. R., IJzerman, H., Urry, H. L., Forscher, P. S., Grahe, J. E., McCarthy, R. J., Musser, E. D., Antfolk, J., Castille, C. M., Evans, T. R., Fiedler, S., Flake, J. K., Forero, D. A., Janssen, S. M. J., Keene, J. R., Protzko,

J., Aczel, B., . . . Chartier, C. R. (2018). The Psychological Science Accelerator: Advancing psychology through a distributed collaborative network [Publisher: SAGE Publications Sage CA: Los Angeles, CA]. *Advances in Methods and Practices in Psychological Science*, *1*, 501–515.

Munafò, M., Nosek, B., Bishop, D., Button, K., Chambers, C., Du Sert, N., Simonsohn, U., Wagenmakers, E.-J., Ware, J., & Ioannidis, J. (2017). A manifesto for reproducible science. *Nature Human Behaviour*, *1*, 0021.

Myung, J., Karabatsos, G., & Iverson, G. J. (2005). A Bayesian approach to testing decision making axioms. *Journal of Mathematical Psychology*, *49*, 205–225.

Nature Publishing Group. (2016). Reality check on reproducibility [editorial]. *Nature*, *533*, 437.

Nosek, B., Alter, G., Banks, G. C., Borsboom, D., Bowman, S. D., Breckler, S. J., Buck, S., Chambers, C. D., Chin, G., Christensen, G., Contestabile, M., Dafoe, A., Eich, E., Freese, J., Glennerster, R., Goroff, D., Green, D. P., Hesse, B., Humphreys, M., . . . Yarkoni, T. (2015). Promoting an open research culture. *Science*, *348*, 1422–1425.

Open Science Collaboration, T. (2015). Estimating the reproducibility of psychological science. *Science*, *349*, aac4716.

Pashler, H., & Harris, C. R. (2012). Is the replicability crisis overblown? Three arguments examined. *Perspectives on Psychological Science*, *7*, 531–536.

Pashler, H., & Wagenmakers, E.-J. (2012). Editors' introduction to the special section on replicability in psychological science: A crisis of confidence? *Perspectives on Psychological Science*, *7*, 528–530.

R Core Team. (2021). *R: A language and environment for statistical computing*. R Foundation for Statistical Computing. Vienna, Austria. https://www.R-project.org/

Rasch, G. (1960). *Probabilistic models for some intelligence and attainment tests*. The Danish Institute for Educational Research.

Regenwetter, M., Cavagnaro, D. R., Popova, A., Guo, Y., Zwilling, C., Lim, S. H., & Stevens, J. R. (2018). Heterogeneity and parsimony in intertemporal choice. *Decision*, *5*, 63–94.

Regenwetter, M., Dana, J., & Davis-Stober, C. P. (2011). Transitivity of preferences. *Psychological Review*, *118*, 42–56.

Rouder, J. N. (2014). Optional stopping: No problem for Bayesians. *Psychonomic Bulletin & Review*, *21*, 301–308.

Rouder, J. N., & Morey, R. D. (2018). Teaching Bayes' theorem: Strength of evidence as predictive accuracy. *The American Statistician*, 1–5.

Sarafoglou, A., Hoogeveen, S., Matzke, D., & Wagenmakers, E.-J. (2020). Teaching good research practices: Protocol of a research master course. *Psychology Learning & Teaching*, *19*, 46–59.

Schmidt, S. (2009). Shall we really do it again? The powerful concept of replication is neglected in the social sciences. *Review of General Psychology*, *13*, 90–100.

Simmons, J., Nelson, L., & Simonsohn, U. (2011). False–positive psychology: Undisclosed flexibility in data collection and analysis allows presenting anything as significant. *Psychological Science*, *22*, 1359–1366.

Simons, D. (Ed.). (2018). Challenges in making data available. *Advances in Methods and Practices in Psychological Science*, *1*.

Stefan, A. M., Gronau, Q. F., Schönbrodt, F. D., & Wagenmakers, E.-J. (2019). A tutorial on Bayes factor design analysis using informed priors. *Behavior Research Methods*, *51*, 1042–1058.

Tijmstra, J., Hoijtink, H., & Sijtsma, K. (2015). Evaluating manifest monotonicity using Bayes factors. *Psychometrika*, *80*, 880–896.

van Doorn, J., Matzke, D., & Wagenmakers, E.-J. (2020). An in-class demonstration of Bayesian inference. *Psychology Learning & Teaching*, *19*, 36–45.

Verhagen, J., & Wagenmakers, E.-J. (2014). Bayesian tests to quantify the result of a replication attempt. *Journal of Experimental Psychology: General*, *143*, 1457–1475.

Wagenmakers, E.-J. (2007). A practical solution to the pervasive problems of p-values. *Psychonomic Bulletin & Review*, *14*, 779–804.

Wagenmakers, E.-J., Lodewyckx, T., Kuriyal, H., & Grasman, R. (2010). Bayesian hypothesis testing for psychologists: A tutorial on the Savage–Dickey method. *Cognitive Psychology*, *60*, 158–189.

Wagenmakers, E.-J., Marsman, M., Jamil, T., Ly, A., Verhagen, J., Love, J., Selker, R., Gronau, Q., Smira, M., Epskamp, S., Matzke, D., Rouder, J., & Morey, R. (2018). Bayesian inference for psychology. Part I: Theoretical advantages and practical ramifications. *Psychonomic Bulletin & Review*, *25*, 35–57.

Wasserstein, R. L., Lazar, N. A., American Statistical Association, Gelman, A., Loken, E., Johnson, V., Leek, J., Morganstein, D., Wasserstein, R., Nuzzo, R., & and others. (2016). ‘‘Editorial”, Basic and Applied Social psychology.

Wicherts, J., Borsboom, D., Kats, J., & Molenaar, D. (2006). The poor availability of psychological research data for reanalysis. *American Psychologist*, *61*, 726–728.

Witmer, J. (2017). Bayes and MCMC for undergraduates. *The American Statistician*, *71*, 259–264.

Wyman, A. J., & Vyse, S. (2008). Science versus the stars: A double-blind test of the validity of the neo five-factor inventory and computer-generated astrological natal charts. *The Journal of General Psychology*, *135*, 287–300.